\documentclass[10pt,letterpaper]{article}
\usepackage{opex3}

\usepackage{amssymb}
\usepackage{bm}% bold math

\usepackage[french]{babel}
\usepackage{amsmath}
\usepackage{amsthm}
\usepackage{amsfonts}
\usepackage{amssymb}
\usepackage{minitoc}
\usepackage{fancyhdr}
\usepackage{esint}
\usepackage{float}
\usepackage{subfig}
\usepackage{wrapfig}

\begin{document}

%%%%%%%%%%%%%%%%%% title page information %%%%%%%%%%%%%%%%%%
\title{Plasmonic space folding: focussing surface plasmons via negative refraction in complementary media}

\author{Muamer Kadic, Sebastien Guenneau, Stefan Enoch}
\address{Institut Fresnel, CNRS, Aix-Marseille
Universit\'e,\\Campus universitaire de Saint-J\'er\^ome,
 13013 Marseille, France}
\email{sebastien.guenneau@fresnel.fr} %% email address is required
\author{S. Anantha Ramakrishna}
\address{Department of Physics, Indian Institute of Technology, Kanpur 208016, India}



\begin{abstract}
We extend designs of perfect lenses to the focussing of
surface plasmon polaritons (SPPs) propagating at the interface between two anisotropic media of opposite permittivity sign. We identify the
role played by the components of anisotropic and heterogeneous tensors of
permittivity and permeability, deduced from a coordinate
transformation, in the dispersion relation governing propagation of
SPPs. We illustrate our theory with three-dimensional finite element computations for focussing of SPPs by
 perfect flat and cylindrical lenses.
\end{abstract}

%\ocis{(240.0240) Optics at surfaces;
%(240.6680) Surface plasmons;
%(160.1190) Materials : Anisotropic optical materials;
%(260.2110) Physical optics : Electromagnetic optics;
%(160.3918) Materials : Metamaterials} % REPLACE WITH CORRECT OCIS CODES FOR YOUR ARTICLE

%%%%%%%%%%%%%%%%%%%%%%% References %%%%%%%%%%%%%%%%%%%%%%%%%

%%%%%%%%%%%%%%%%%%%%%%%%%%  body  %%%%%%%%%%%%%%%%%%%%%%%%%%

\section{Introduction}
Focusing light using negative refraction is an emerging subject mixing fascinating and elusive features.
It is now well known that one can reverse the flow of light with negative refractive index materials, within which
light takes the ``{\it wrong}'' turn in accordance with inverted Snell-Descartes laws of refraction \cite{veselago,jp2000,ramakrishna}.
However, there has been a growing interest in a better control of light through transformational
optics, following the recent proposals by Pendry et al. \cite{pendry} and Leonhardt
\cite{leonhardt06}. The former seminal paper demonstrates the possibility of designing
a cloak that  renders any object inside it invisible to
electromagnetic radiation (using the covariant structure of Maxwell's equations),
while the latter concentrates on the ray optics limit (using conformal mappings in the
complex plane). In both cases, the cloak consists of a meta-material
whose physical properties (permittivity and permeability) are spatially varying as well as anisotropic.

Importantly, if one considers a point source located inside the cloak,
it appears to radiate from a shifted location: Cloaking is nothing but a mirage effect \cite{zolla}.
If one adopts this viewpoint, complementary media \cite{sar1,sar2} within which focussing
effects via negative refraction occur can be designed using geometric transforms by mapping the image plane
back onto the source plane \cite{ulfprog,sar3}. Such lenses were coined as Alice mirrors in
\cite{alice} since they exhibit certain anti-symmetric features. These make also
possible perfect and poor man's cylindrical lenses \cite{perf2,milton2}, as well as generalized
corner and checkerboard lenses \cite{sar2,sebjohn,sar4,sar5}.

However, there are other types of electromagnetic waves well worth controlling such as surface plasmon polaritons
(SPPs) that allow for fascinating applications in the emerging field of nano-optics \cite{maradudin,maier}.
Back in 1998, Ebbesen et al. established that resonant excitation of
SPPs on a metal film perforated at subwavelength periodicities leads to unusually high transmission
coefficients~\cite{ebbesen}.
Pendry, Martin-Moreno and Garcia-Vidal further showed in 2004 that one can
manipulate SPPs by structuring perfectly conducting surfaces in a periodic manner \cite{science2004}. 
Other plasmonic metamaterials include plasmonic shells with a suitable out-of-phase polarizability in order to
compensate the scattering from the knowledge of the electromagnetic
parameters of the object to hide, and external cloaking, whereby a
plasmonic resonance cancels the external field at the location of a
set of electric dipoles \cite{milton2,engheta,javier,baumeier}.

In the present paper, we extend the design of transformation based
perfect lens to the focussing of surface plasmon polaritons (SPPs). The
field of plasmonics has matured to the point where an introduction to
the optical properties of metals and sometimes even surface plasmons
is included in many standard texts in solid-state physics and
optics, with a state of the art on meta-surfaces available in \cite{ramakrishna,maier}.
Interestingly, Agranovich et al. have shown in 2004 that one can obtain negative refraction 
for exciton-plasmon waves in organic and gyrotropic material using a 
surface transition layer \cite{agranovich}.

Our main contribution here is that we explain how one can manipulate SPPs in such a
way that the electromagnetic space where they live is folded back onto itself, so that
we can extend the proposal of Pendry and Ramakrishna to build generalized lenses using the
original concept of complementary media \cite{sar1,sar2}
to the area of plasmonic waves. This allows for instance the design of perfect corners,
corner lenses and the like \cite{sar2,sebjohn,sar4,sar5}, using transformational plasmonics tools
\cite{muamer,rengprl,zangnanoletter,garcananoletter}.
Importantly, we also perform full wave computations
to exemplify the imaging effects of SPPs through negative refraction in a novel class of generalised plasmonic
lenses. 

\section{Transformational plasmonics and generalized perfect lenses in Cartesian coordinates for SPPs}

Let us consider two semi-infinite regions separated by a plane interface at $x_2=0$.
%\ref{fig1}.
The upper region ($x_2>0$) is filled with air i.e. with relative permittivity $\varepsilon_1=1$
(resp. relative permeability $\mu_1=1$), while the lower region ($x_2<0$) is filled with a Drude metal i.e.
with relative permittivity
\begin{equation}
\varepsilon_2=1-\frac{\omega_p^2}{\omega^2+\i\gamma\omega}
\label{drude}
\end{equation}
 (resp. relative permeability $\mu_2=1$): here, some gold with the plasma frequency ($\omega_p=2175$ THz) and characteristic collision frequency ($\gamma=4.35$ THz).
We would like to map these two isotropic homogeneous media on two metamaterials described by anisotropic heterogeneous
matrices of permittivity and permeability given by \cite{zolla}
\begin{equation}
\varepsilon' = \varepsilon{\bf T}^{-1} , \mbox{ and } \mu' = \mu{\bf T}^{-1} \; .
\label{equage}
\end{equation}
where ${\bf T}={\bf J}^T {\bf J}/\hbox{det}({\bf J})$ is the transformation matrix constructed using the Jacobian associated
with the change of coordinates. Let us emphasize here that in our numerical implementation, we make use of finite edge elements
that are nothing but discrete Whitney differential forms, and behave nicely under pull-back transforms. Thus, in what
follows, we always map the destination domain onto the original one (so we consider the inverse transforms).

\begin{figure}[H]
\begin{center}
\includegraphics[scale=0.3]{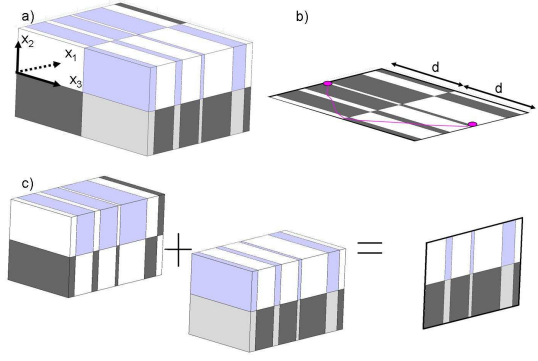}
\caption{Principle of plasmonic space folding within complementary media: (a) An alternation of positively
and negatively refracting index media in all three space dimensions cancels the plasmonic path (note that
the bottom part (in grey color) is filled with Drude complementary media; (b) Typical propagation path for a SPP
at the $x_2=0$ interface; (c) Underlying mechanism.}
\label{fig1}
\end{center}
\end{figure}

\noindent The original perfect lens
presupposed a slab of material with $\epsilon$ = -1 and $\mu$ = -1. Let us now derive the
result using the powerful tool of transformational plasmonics \cite{muamer}.
In order to design generalized lenses for surface plasmon polaritons,
we want to fold the plasmonic space back onto itself, and this leads to negative
coefficients within the permittivity and permeability matrices. The
coordinate transformation is given by
\begin{eqnarray}
%\left\lbrace
%\begin{array}{ll}
x'_1 & =& x_1  , \nonumber \\
x'_2 & = &x_2 , \nonumber \\
x'_3 & = &x_3 -d  , ~~\mathrm{ if } ~x'_3 < d/2 , \nonumber \\
 ~&~&\mathrm{or } ~-x_3 ~~\mathrm{if } ~-d/2 < x_3 < d/2, \nonumber \\
  ~&~&\mathrm{or } ~x_3 +d ~~\mathrm{if}~ d/2 < x_3, \nonumber
%\end{array}
\label{transfolens1d}
\end{eqnarray}
where $d$ is the thickness of the generalized lens.

The above coordinate transform
%(\ref{transfolens1d})
leads to the identity
for the transformation matrix ${\bf T}$ outside the lens, whereas
inside the lens i.e. for $-d/2 < x'_3 < d/2$, $\partial x_3/\partial
x'_3=-1$ which flips the sign of ${T}_{33}$, so that the material
properties differ from free space only for points along the $x_3=x'_3$ direction.
%whereby the scalar permittivity $\varepsilon=\varepsilon_1$ and the
%transformed (scalar) permittivity $\varepsilon'=\varepsilon_2$, and
%similarly for the permeability.

\begin{figure}[H]
\begin{center}
\includegraphics[scale=0.3]{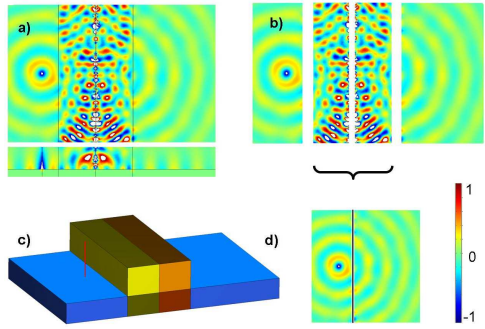}
\caption{SPP focussing in a checkerboard lens: (a) Top and side views of the computed field for a SPP line source;
(c) Four complementary media at $700$ nanometers:
$\varepsilon_1^1=\mu_1^1=2$ in yellow block, $\varepsilon_2^1=\mu_2^1=-2$
in orange block with their Drude metal counterparts $\varepsilon_1^2=\mu_1^2=2(1-\frac{\omega_p^2}{\omega^2+\i\gamma\omega})$ in dark yellow block, $\varepsilon_2^2=\mu_2^2=-2(1-\frac{\omega_p^2}{\omega^2+\i\gamma\omega})$
in red block underneath.
(b) The middle regions cancel each other, leading to (d). 
}
\label{fig2}
\end{center}
\end{figure}

\begin{figure}[H]
\begin{center}
\includegraphics[scale=0.3]{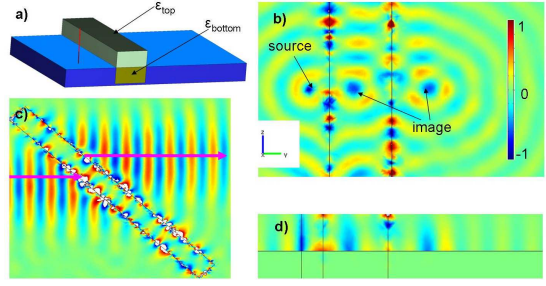}
\caption{Perfect plasmonic lens via negative refraction at $700$ nanometers:
(a) The medium $\varepsilon_1^1=\mu_1^1=1$ in the yellow block
and its Drude metal counterpart underneath $\varepsilon_1^2=1-\frac{\omega_p^2}{\omega^2+\i\gamma\omega}$, $\mu_1^2=1$;
(b) Top and side views of the
computed field for a SPP line source; (c) Negatively refracted SPP beam making
an angle of $45$ degrees with the lens.}
\label{fig3}
\end{center}
\end{figure}

The above analysis demonstrates that SPP focussing will always occur
irrespective of  the medium in which the lens is embedded. This is
true for any medium which is mirror antisymmetric about a vertical
plane. In the plasmonic case, we can define two complementary media
above the metal surface and two complementary media within the metal as
\begin{equation}
\begin{array}{ll}
        \epsilon^1_j &= + \varepsilon_j\epsilon(x_1,x_2), \mu^1_j = +\mu(x_1,x_2), -d < x_3 <
        0 \; , \nonumber
        \\
\epsilon^2_j &= - \varepsilon_j\epsilon(x_1,x_2), \mu^2_j = -\mu(x_1,x_2),  0 < x_3 < d \; .
\end{array}
\end{equation}
Case $j=1$ corresponds to the top region: $0< x_2 <+\infty \;$, while case $j=2$ corresponds to the bottom region  $-\infty< x_2 <0 \;$,
%where $i=1$ for $x_2>0$ and $i=2$ for $x_2<0$
with $\varepsilon_1=1$ and $\varepsilon_2$ as in (\ref{drude}),
%will show identical focussing.

We report in Figure \ref{fig2} some finite element computations showing focussing of a SPP at a wavelength of $700$ nm
for two such complementary media i.e. a checkerboard lens
for the special case $\varepsilon(x_1,x_2)=\mu(x_1,x_2)=2$. The cancellation of {\it plasmonic} space is noted.
We further show in Figure \ref{fig3} a focussing effect through a plasmonic flat lens with $\varepsilon(x_1,x_2)=\mu(x_1,x_2)=1$.
The source and the perfect images indeed satisfy the inverted Snell-Descartes laws of negative refraction, as
shown for a SPP beam incident upon the lens at an angle of 45 degrees. 

However, in the case of anisotropic
$\underline{\underline{\varepsilon}^i_j}$ and
$\underline{\underline{\mu}^i_j}$, the transformed media are now
characterized in the region $-d < x_3 <0 $ by
%(\ref{equage}):

\begin{equation}
\begin{array}{lll}
&\underline{\underline{\varepsilon^{'i}_j}} 
= \varepsilon_j(-1)\left(
\begin{array}{rrr}
        1 & 0 & 0 \\
        0 & 1 & 0 \\
        0 & 0 &-1
         \end{array} \right)
\left( \begin{array}{lll}
        \varepsilon_{11}^i & \varepsilon_{12}^i & \varepsilon_{13}^i \\
        \varepsilon_{21}^i & \varepsilon_{22}^i & \varepsilon_{23}^i \\
        \varepsilon_{31}^i & \varepsilon_{32}^i & \varepsilon_{33}^i
         \end{array} \right)
\left(
\begin{array}{rrr}
        1 & 0 & 0 \\
        0 & 1 & 0 \\
        0 & 0 & -1
         \end{array} \right)
  %      = \left( \begin{array}{lll}
   %     -\varepsilon_{11}^i & -\varepsilon_{12}^i & +\varepsilon_{13}^i \\
    %    -\varepsilon_{21}^i & -\varepsilon_{22}^i & +\varepsilon_{23}^i \\
     %   +\varepsilon_{31}^i & +\varepsilon_{32}^i & -\varepsilon_{33}^i
      %   \end{array} \right),
         \\
&\underline{\underline{\mu^{'i}_j}} 
%=\rm{det}(\mathbf{J})
%\left(\mathbf{J}^{-1}\underline{\underline{\mu}}\mathbf{J}^{-T}\right)
= (-1)\left(
\begin{array}{rrr}
        1 & 0 & 0 \\
        0 & 1 & 0 \\
        0 & 0 &-1
         \end{array} \right)
\left( \begin{array}{lll}
        \mu_{11}^i & \mu_{12}^i & \mu_{13}^i \\
        \mu_{21}^i & \mu_{22}^i & \mu_{23}^i \\
        \mu_{31}^i & \mu_{32}^i & \mu_{33}^i
         \end{array} \right)
\left(
\begin{array}{rrr}
        1 & 0 & 0 \\
        0 & 1 & 0 \\
        0 & 0 & -1
         \end{array} \right)
 %        =
%\left( \begin{array}{lll}
 %       -\mu_{11}^i & -\mu_{12}^i & +\mu_{13}^i \\
  %      -\mu_{21}^i & -\mu_{22}^i & +\mu_{23}^i \\
   %     +\mu_{31}^i & +\mu_{32}^i & -\mu_{33}^i
    %     \end{array} \right),
\end{array}
\end{equation}
%which is in accordance with (\ref{sarlens2}).

We note that there is no change in the impedance of the media, since
the permittivity and permeability undergo the same geometric
transformation: the perfect lens is impedance-matched with its
surrounding medium (air, say) so that no reflection will occur at
its interfaces.

%Altogether, if we consider a complex medium described by general
%dielectric permittivity and magnetic permeability tensors given by...........

Then the resulting complementary medium (see Figure \ref{fig1})
%\ref{fig1}
is given by
\begin{equation}
\underline{\underline{\varepsilon^{'i}_j}} = \varepsilon_j\left( \begin{array}{lll}
       -\varepsilon_{11}^i & -\varepsilon_{12}^i & +\varepsilon_{13}^i \\
        -\varepsilon_{21}^i & -\varepsilon_{22}^i & +\varepsilon_{23}^i \\
        +\varepsilon_{31}^i & +\varepsilon_{32}^i & -\varepsilon_{33}^i
         \end{array} \right), ~~~
\underline{\underline{\mu^{'i}_j}} = \left( \begin{array}{lll}
       -\mu_{11}^i & -\mu_{12}^i & +\mu_{13}^i \\
        -\mu_{21}^i & -\mu_{22}^i & +\mu_{23}^i \\
        +\mu_{31}^i & +\mu_{32}^i & -\mu_{33}^i
         \end{array} \right), \;  0 < x_3 < d\; ,
\label{sarlens2}
\end{equation}
where $\varepsilon_1=1$ and $\varepsilon_2$ are as in (\ref{drude}), and $j=1$ for
$x_2>0$ and $j=2$ for $x<0$. 
This is a generalization to transformational plasmonics of the result first derived in \cite{sar2} and
retrieved again using group theory (symmetries of Maxwell's equations) in
\cite{sar4}. The entries in
$\underline{\underline{\varepsilon^i_j}}$ and
$\underline{\underline{\mu^i_j}}$ can also be spatially varying along
$x_1$ and $x_2$. This covers the case of perfect corner reflectors of
$2n$-fold skew-symmetry and we are therefore ensured of the
cancellation of the plasmonic path. It is worth noting that the
generalized lens theorem was also applied to infinite checkerboards
of skew-symmetry in \cite{sar4,sar5}.

We now extend the dispersion relation for a surface plasmon at the
interface between a transformed medium and a transformed metal  \cite{muamer} to the case where we have
four media described by diagonal
tensors of relative permittivity and permeability:
$\underline{\underline{\varepsilon^i_j}}=\hbox{diag}(\varepsilon_{11,j}^i,\varepsilon_{22,j}^i,\varepsilon_{33,j}^i)$
and
$\underline{\underline{\mu^i_j}}=\hbox{diag}(\mu_{11,j}^i,\mu_{22,j}^i,\mu_{33,j}^i)$
with $j=1$ when $x_2>0$ and $j=2$ when $x_2<0$, see Figure \ref{fig1} (a),
%\ref{fig1}(a),
The SPP propagating at the metal surface has a wavenumber satisfying:

\begin{equation}
k_3^i=\frac{\omega}{c}\sqrt{\frac{\varepsilon_{33,2}^i\varepsilon_{33,1}^i
(\mu_{22,2}^i\varepsilon_{11,1}^i-\mu_{22,1}^i\varepsilon_{11,2}^i)}
{\varepsilon_{11,1}^i\varepsilon_{33,1}^i-\varepsilon_{11,2}^i\varepsilon_{33,2}^i}}
\; . \label{spp1newb}
\end{equation}
where $i=1,2$ denotes the $i^{th}$ transformed medium and its Drude metal counterpart. This formula
also characterizes the wavenumber of an SPP propagating at a multi-layered metal surface, whereby
the number $i$ of complementary regions is only constrained by the overall
propagation length of the SPP.  

\begin{figure}[H]
\begin{center}
\includegraphics[scale=0.3]{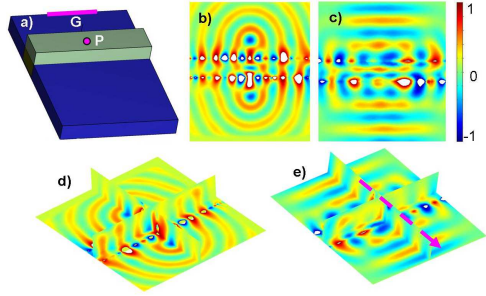}
\caption{Compact anisotropic plasmonic lens at 700 nanometers: (a) 3d view of a three-dimensional anisotropic perfect lens;
two cases have been studied: a Gaussian beam of waste $2100$ nanometers on normal incidence launched from the line G and a line source at point P; 
b)Magnetic field phase for a line source placed at the point P. 
c)Magnetic field phase for a Gaussian beam launched from the line G.
d) 3d view of b)
e) 3d view of c).}
\label{anisotropic}
\end{center}

\end{figure}

Following the work by Shen et al. \cite{kafesaki} on two-dimensional anisotropic perfect lenses,
we report in Figure \ref{anisotropic} some computations for an SPP (Gaussian beam) incident
from the left on such a lens in the three-dimensional transformational plasmonic setting. Above the metal interface, the
transformed medium is described by
$\underline{\underline{\varepsilon_1}}=\hbox{diag}(\varepsilon_{11,1},\varepsilon_{22,1},\varepsilon_{33,1})=
(-2-0.001\i,-2-0.001\i,-0.5-0.001\i)$
and
$\underline{\underline{\mu_1}}=\hbox{diag}(\mu_{11,1},\mu_{22,1},\mu_{33,1})=(-2-0.001\i,-2-0.001\i,-0.5-0.001\i)$
for the bottom layer, the transformed metal is described by:
$\underline{\underline{\varepsilon_2}}=\hbox{diag}(\varepsilon_{11,2},\varepsilon_{22,2},\varepsilon_{33,2})=\varepsilon_2
(-2-0.001\i,-2-0.001\i,-0.5-0.001\i)$
and
$\underline{\underline{\mu_2}}=\hbox{diag}(\mu_{11,2},\mu_{22,2},\mu_{33,2})=(-2-0.001\i,-2-0.001\i,-0.5-0.001\i)$
which is a particular case of (\ref{sarlens2}). In panels (b) and (d) of Figure \ref{anisotropic}, we place a line source
within the slab lens (shown by point $P$ in panel (a)), and we observe two perfect images on either sides of the anisotropic lens, in accordance with the
inverted Snell-Descartes laws of refraction. We note that on the vertical sides of this lens, the following condition is met:

\begin{equation}
\frac{\varepsilon_{11,j}}{\mu_{22,j}}=\frac{\varepsilon_j}{\mu_j} \; , \; \varepsilon_{11,j}\varepsilon_{33,j}=(\varepsilon_j)^2 \; ,
\frac{\mu_{11,j}}{\varepsilon_{22,j}}=\frac{\mu_j}{\varepsilon_j} \; , \; \mu_{11,j}\mu_{33,j}=(\mu_j)^2 \; ,
\end{equation}
where $\varepsilon_1=\mu_1=1$ describe the ambient medium above the surface (air) and $\varepsilon_2$ is as in (\ref{drude}) and $\mu_2=1$.
Hence, the slab lens is perfectly matched to the surrounding medium.

We numerically check this fact by sending a Gaussian SPP beam with a waste of $3\lambda=2100$ nanometers on this less than usual
lens. The wavefront shrinks within the lens, but forward and backward scatterings do not sense the presence of the lens: the lens is indeed
invisible for any SPP propagating on the metal surface.

\section{Transformational plasmonics and generalized perfect lenses in cylindrical coordinates for SPPs}
\begin{figure}[H]
\begin{center}
\includegraphics[scale=0.3]{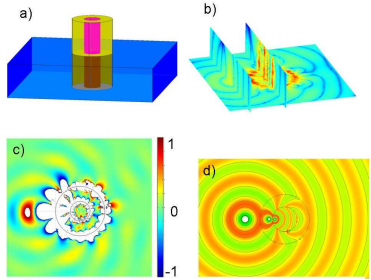}
\caption{Focussing through a cylindrical plasmonic lens for a SPP line source at wavelength $700$ nanometers:
(a) Schematic diagram of the cylindrical lens (the pink region is filled with spatially varying permittivity
${(R_2/R_1)}^4$ above and its Drude metal counterpart below, the gold region is filled with negative
permittivity $-R_2^2/r^2$ and negative permittivity $-1$ above and its Drude metal counterpart below; (b) 3D
plot of the magnitude of the field; (c) 2D computation for TE waves (with parameters of the top regions of a);
(d) Top view of (b).}
\label{fig4}
\end{center}
\end{figure}

Building on the previous perfect lens theory, we can use conformal transformations to bend
the shape of the perfect lens into other geometries such as two concentric cylinders, see Figure \ref{fig4}(a).
The objective is to follow Pendry's proposal of a magnifying glass, and for this purpose the lens must be curved.
The conformal transformation $z' = \ln z$, where $z = x + iy$ is known to preserve the solutions of Laplace's
equation and leaves the values of $\varepsilon$ and $\mu$ unchanged. However, when one departs from the static case,
the previous geometric transform leads to anisotropic heterogeneous permittivity and permeability tensors.

Following the same algorithm as in the previous section we find that a possible design for
a cylindrical perfect lens for SPPs (avoiding spatially varying media outside the lens) is given by 

\begin{equation}
\begin{array}{ccc}
\varepsilon^1_j = \varepsilon_j {(R_2/R_1)}^4 \; , \;& \mu^2_j=1 \; , \; & r\leq R_1 \\
\varepsilon^2_j =- \varepsilon_j R_2^2/r^2 \; , \; & \mu^2_j=-1 \; , \; & R_1\leq r \leq R_2
\end{array}
\end{equation}
with $j=1$ for $x_2>0$, $j=2$ for $x_2<0$, and $\varepsilon_1=1$ and $\varepsilon_2$ as in (\ref{drude}).

We show in panels (b) and (d) of Figure \ref{fig4} some 3D finite element computations
for a SPP line source at a wavelength of  $700$ nm, located inside the central region of the cylindrical lens.
We also depict in panel (c) a 2D numerical simulation for comparison with the case of transverse electric waves
(whereby $\varepsilon_1=\varepsilon_2=1$). We note that in such a configuration, the image is magnified.

\section{Conclusion}
In conclusion, we have studied
analytically and numerically the extension of optical space folding to the area of plasmonics.
This requires anisotropic heterogeneous complementary media deduced from geometric transformations.
Focusing SPP through a markedly enhanced control of their wave trajectories has been demonstrated.
Importantly, our analysis applies mutatis mutandis
for a s-polarized SPP (interchanging the roles of permittivity and permeability).
For illustrative purpose, checkerboard and cylindrical designs of SPP lenses have been proposed and
implemented in the finite element commercial  package COMSOL.
%$\textcircled{R}$.
%These waves obey the Maxwell equations at a flat interface and are
%evanescent in the transverse direction, so that, the problem we have
%treated is somewhat isomorphic to the case of linear surface water
%waves which satisfy a similar dispersion relation to SPPs.

\medskip

The authors are grateful for insightful comments by Mr G. Dupont and
Prof. R. Quidant. MK acknowledges funding from the French ministry
of Higher-Education and Research.  SG, SAR and SE are thankful for
funding to the Indo-French Centre for the Promotion of
Advanced Research, New Delhi under project no. IFCPAR / 3804 -2.

\end{document}